\documentclass[aps,prl,showpacs,superscriptaddress,amsmath,amssymb,floatfix,twocolumn]{revtex4-1}
\usepackage{graphicx}
\usepackage{dcolumn}
\usepackage{bm}
\usepackage{bigstrut}
\begin{document}

\title{Wetting, Algebraic Curves and Conformal Invariance}

\author{A.O.\ Parry}
\affiliation{Department of Mathematics, Imperial College London, London SW7 2BZ, UK}

\author{C.\ Rasc\'{o}n}
\affiliation{GISC, Departamento de Matem\'aticas, Universidad Carlos III de Madrid, 28911 Legan\'es, Madrid, Spain}

\begin{abstract}
Recent studies of wetting in a two-component square-gradient model of interfaces in a fluid mixture, showing three-phase bulk coexistence, have revealed some highly surprising features. Numerical results show that the density profile paths, which form a tricuspid shape in the density plane, have curious geometric properties, while conjectures for the analytical form of the surface tensions imply that nonwetting may persist up to the critical end points, contrary to the usual expectation of critical point wetting. Here, we solve the model exactly and show that the profile paths are conformally invariant quartic algebraic curves that change genus at the wetting transition. Being harmonic, the profile paths can be represented by an analytic function in the complex plane which then conformally maps the paths onto straight lines. Using this, we derive the conjectured form of the surface tensions and explain the geometrical properties of the tricuspid and its relation to the Neumann triangle for the contact angles. The exact solution confirms that critical point wetting is absent in this square-gradient model.
\end{abstract}



\maketitle

\begin{figure*}[t]
\includegraphics[width=1.8\columnwidth]{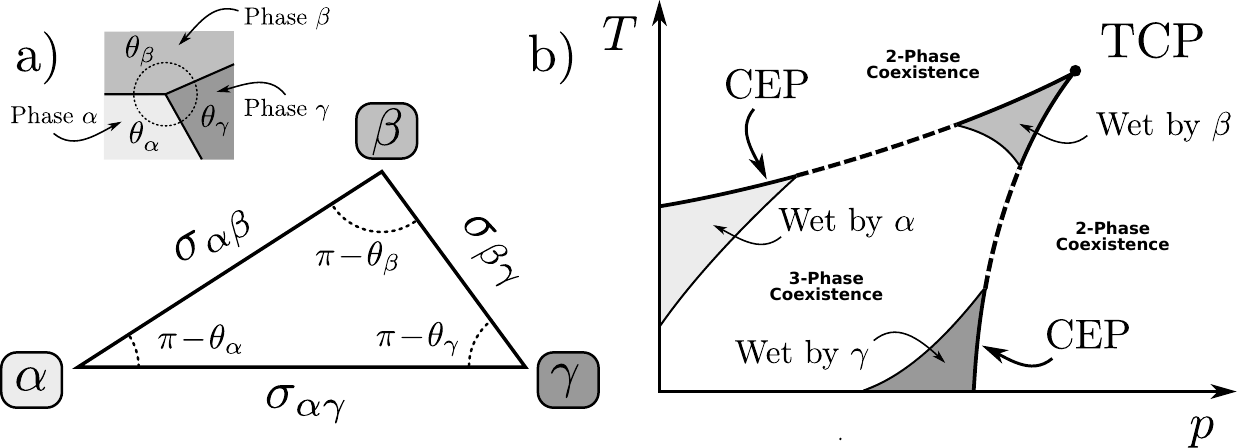}
\caption{\label{Fig1} 
a) Neumann triangle for the surface tensions and contact angles (inset), $\theta_\alpha$, $\theta_\beta$ and $\theta_\gamma$ and b) Indekeu-Koga surface phase diagram showing the regions of complete wetting by $\alpha$, $\beta$ or $\gamma$ in the three phase region of the temperature $T$ and pressure $p$ plane. Nonwetting persists up to the dashed sections of the lines of critical end points (CEP) which terminate at a bulk tricritical point (TCP).}
\end{figure*}

At a wetting transition, the adsorption of a fluid phase intruding between two other phases changes from microscopic to macroscopic, associated with the vanishing of a contact angle – for reviews see \cite{Sullivan1986,Dietrich1988,Schick1990,Forgacs1991,Bonn2009}. Wetting transitions were first predicted by Cahn \cite{Cahn1977} and Ebner and Saam \cite{Ebner1977}, using model density functional theory descriptions of a solid-fluid interface \cite{Evans1979}, with Cahn also offering a simple argument that such a transition must occur prior to the bulk critical temperature – hence the terminology critical point wetting (CPW). While the Cahn argument is flawed CPW is always found in models where the solid-fluid and fluid-fluid forces are both short-ranged or both long-ranged \cite{Sullivan1979,Sullivan1981,Abraham1980,Nakanishi1982,Parry2023b} and is only supressed when there is a mismatch between their ranges \cite{DeGennes1983,Nightingale1985,Evans2019,Parry2024}. For a mixture of three fluid phases, $\alpha$, $\beta$ and $\gamma$, however, the necessity of CPW has remained unclear, despite there being many more experiments \cite{Moldover1980,Bonn2001}. In this scenario, there are three contact angles $\theta_\mu$ ($\mu=\alpha, \beta, \gamma$) that satisfy the Neumann triangle involving the surface tensions -- see Fig.\ \ref{Fig1} a). In a recent article, Indekeu and Koga \cite{Indekeu2022} have argued that, even in a model system with short-ranged forces, CPW need not occur. Their analysis was based on an elegant mean-field square-gradient model first introduced by Koga and Widom for studying wetting in the region of three phase fluid coexistence bounded by two critical end point lines meeting at a tricritical point \cite{Koga2008,Koga2016}.
They provided numerical evidence that a continuous wetting transition occurs just before the bulk tricritical point, and guessed, without any derivation, analytical expressions for the surface tensions along a line of symmetry.
This was generalised further by Koga and Indekeu \cite{Koga2019} to the whole phase diagram and predicted that the lines of wetting transitions do not converge at the critical end points, leaving instead a nonwetting gap \cite{Indekeu2022} – see Fig.\ \ref{Fig1} b).
This is an extremely interesting proposal, which challenges not only the usual CPW scenario, but also the apparently reasonable assumption that the interfacial behaviours near a rigid wall and near a noncritical fluid should be similar \cite{Fisher1990a,Fisher1990b}. Here we show that the square-gradient model, hereafter the Koga-Widom-Indekeu (KWI) model, is exactly solvable, deriving the wetting phase boundaries and the density profile paths analytically, which we show are conformally invariant quartic algebraic curves. We then use the conformal invariance to derive the Koga-Indekeu conjecture for the surface tension and further show that the tricuspid shape, defined by the profile paths, is a curvilinear representation of the Neumann triangle for the contact angles. 

The KWI model is a two-component square-gradient theory of wetting in a mixture of three fluid phases which generalises the much simpler one-component theory where one phase ($\beta$ say) always wets the $\alpha\gamma$ interface \cite{Rowlinson1982}. The grand potential functional per unit area is written
\begin{equation}
\Omega[\rho_1,\rho_2]=\int_{-\infty}^\infty d z\;\left[\,\frac{1}{2}\sum_{i=1,2}\rho'_i (z)^2\;+\;\omega\,(\rho_1,\rho_2)\right]
\label{Omega}
\end{equation}
where $\rho_1(z)$ and $\rho_2(z)$ are the two density profiles, the prime denotes differentiation w.r.t.\ the co-ordinate $z$, normal to the plane of the interfaces, and
\begin{equation}
\omega(\rho_1,\rho_2)=\prod_\mu \left[\,(\rho_1-\rho_1^\mu)^2+(\rho_2-\rho_2^\mu)^2\,\right]
\end{equation}
for $\mu=\alpha,\beta,\gamma$, which is a sixth order polynomial. 
The bulk densities are related to the temperature $t$ and pressure $s$ scaling variables via $(\rho_1^\mu)^3-3t\rho_1^\mu+2s=0$ with $\rho_2^\mu=-(\rho_1^\mu)^2$
\cite{Indekeu2022,Rowlinson1982,Griffiths1974}. The bulk phase diagram has critical end point lines along $st^{-3/2}=\pm 1$ which converge at the bulk tricritical point at $t=s=0$. 
Here, we leave the bulk densitiues as arbitrary but ordered, $\,\rho_1^\alpha\le \rho_1^\beta \le \rho_1^\gamma$, and note only that they form the vertices of a triangle in the density plane. Minimization of $\Omega$ leads to a pair of Euler-Lagrange equations $\;\rho_i^{\prime\prime}(z)=\partial_{\rho_i}\omega(\rho_1,\rho_2)\,$ ($i=1,2$), whose solution, subject to bulk boundary conditions at $z=\pm\infty$, determines the equilibrium density profiles $\rho_1(z)$ and $\rho_2(z)$ for the three possible $\alpha\beta$, $\beta\gamma$ and $\alpha\gamma$ interfaces. These density profiles can be represented by a path $\rho_2(\rho_1)$ connecting the bulk densities in the $(\rho_1,\rho_2)$ plane which, for partial wetting, define a curvilinear tricuspid with internal vertex angles $\widetilde\alpha$, $\widetilde\beta$ and $\widetilde\gamma$, which we shall determine. Integration of the Euler-Lagrange equations leads to
$\,\frac{1}{2}\left(\rho_1'(z)^2+\rho_2'(z)^2\right)=\omega\left(\rho_1,\rho_2\right)$,
which is equivalent to the conservation of energy in the well-known mechanical analogy \cite{Rowlinson1982}, where the two densities are likened to the co-ordinates of a particle moving in the 2D plane, with total energy $\,E=0\,$ that takes an infinite time to traverse between two peaks of the potential.
The equilibrium value of $\,\Omega\,$
determines the surface tension of the $\alpha\gamma$ interface (say) as
\begin{equation}
\sigma_{\alpha\gamma}=\sqrt{2}\int_{\rho_1^\alpha}^{\rho_1^\gamma} d\rho_1 \sqrt{\,(1+\dot{\rho}_2^{\,2}\,)\;\omega\big(\rho_1,\rho_2(\rho_1)\big)}
\label{tension1}
\end{equation}
where $\,\dot\rho_2=d\rho_2/d\rho_1$ and we have substituted for the function $\rho_2(\rho_1)$ representing the $\alpha\gamma$ path.

\begin{figure*}[t]
\includegraphics[width=1\columnwidth]{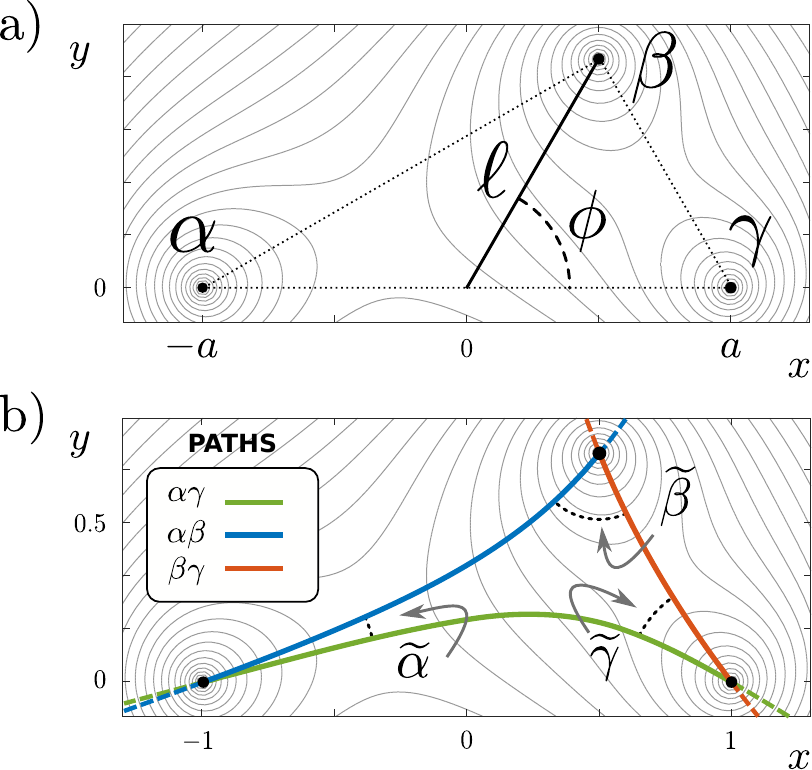}
\caption{\label{Fig2} 
a) Rotated coordinates in the density plane used to define the trajectory $y(x)$ of the $\alpha\gamma$ interface, and the contours of the potential $\omega(x,y)$ (in gray). The bulk phases $\alpha$, $\beta$ and $\gamma$ are located at $(-a,0)$, $(\ell\cos\phi,\ell\sin\phi)$ and $(a,0)$, respectively.
b) A tricuspid in the density plane, together with its interior angles $\widetilde\alpha$, $\widetilde\beta$ and $\widetilde\gamma$, formed from the trajectories of the $\alpha\gamma$, $\alpha\beta$ and $\beta\gamma$ interfaces for a case of partial wetting ($\ell=a=1$, $\phi=\pi/3$). The tricuspid is internal to the triangle connecting the bulk densities.}
\end{figure*}

We now explain how the KWI model may be solved exactly.
First, we introduce new coordinates $(x,y)$ via $\rho_1=Ax+By+(\rho_1^\alpha+\rho_1^\gamma)/2$ and $\rho_2=-Bx+Ay+(\rho_2^\alpha+\rho_2^\gamma)/2$ with $A=(\rho_1^\gamma-\rho_1^\alpha)/2a$ and $B=(\rho_2^\alpha-\rho_2^\gamma)/2a$. This amounts to a simple rotation of the density plane so that we can consider the trajectory $y(x)$ of any one of the interfaces (say the $\alpha\gamma$ path) using a convenient coordinate system in which the bulk densities for $\alpha$, $\beta$ and $\gamma$ appear at $(-a,0)$, $(\ell\cos\phi,\ell\sin\phi)$ and $(a,0)$, respectively. All trajectories can be determined this way with the two other paths obtained using the lengths $a$, $\ell$ and the geometrical median angle $\phi$ appropriate for that side of the triangle, defined by the bulk vertices -- see Fig.\ \ref{Fig2}.
From the Euler-Lagrange equations and the conservation of energy, it follows that the profile path $y(x)$ satisfies 
\begin{equation}
\frac{\ddot{y}}{1+\dot y^2}\;=\;\sum_\mu\;\frac{(y-y_\mu)-\dot y\,(x-x_\mu)}{\,(x-x_\mu)^2+(y-y_\mu)^2\;}
\label{curve}
\end{equation}
where the dot denotes differentiation w.r.t.\ $x$ and the $(x_\mu,y_\mu)$ are the locations of the bulk vertices of the triangle at $\alpha$, $\beta$ and $\gamma$. The $\alpha\gamma$ path begins at $(-a,0)$ and ends at $(a,0)$, and does not pass through $(\ell\cos\phi,\ell\sin\phi)$ provided the interface is not wet by $\beta$. The three terms on the RHS of (\ref{curve}) are homogeneous in $(y-y_\mu)/(x-x_\mu)$ and the equation may be integrated leading to
\begin{equation}
\tan^{-1}\dot y\;\,+\;\sum_\mu\tan^{-1}\left(\frac{\,y-y_\mu}{\,x-x_\mu}\right)\;=\;\phi
\label{ydot}
\end{equation}
implying a local conservation of angles between the path and the vertices. We note that the trajectory can be continued through the bulk densities at $\alpha$ and $\gamma$, where the asymptotes are straight lines separated by the required scattering angle of $3\pi/4$. Next, we combine 
the inverse tangents to show that the first derivative satisfies $\dot y(x) = - P(x,y)/Q(x,y)$ where $P$ and $Q$ are cubic polynomials in $x$ and $y$ satisfying $ \partial_y P= \partial_x Q$, which implies that the equation is an exact differential of a function $u(x,y)$ with $\partial_x u=P$ and $\partial_y u=Q$. Integrating and applying the boundary conditions that the trajectory passes through $(\pm a,0)$, gives $u(x,y)=u_{\alpha\gamma}$ where
\begin{equation}
\begin{split}
u(x,y) & =\cos\phi\,\left(x^3y-xy^3-a^2xy\right)+\ell y\,\left(\frac{y^2}{3}-x^2+a^2\right) \\[.2cm]
 & +\sin\phi\;\left(\frac{3}{2}\,x^2y^2-\frac{x^4+y^4}{4}+\frac{a^2}{2}(x^2-y^2)\right)
 \end{split}
\label{path}
\end{equation}
and $u_{\alpha\gamma}=a^4\sin\phi/4$.
This is the analytical expression for the general $\alpha\gamma$ path, provided it is not wet by $\beta$. The path is unique, implying that any wetting transition is continuous.

From the first integral (\ref{ydot}), it also follows that the sum of the interior angles of the tricuspid shape satisfy
\begin{equation}
\widetilde\alpha+\widetilde\beta+\widetilde\gamma=\frac{\pi}{2}
\label{angles}
\end{equation}
as conjectured by Koga and Widom from numerical results \cite{Koga2008,Koga2016}. The full significance of this relation will emerge later. However it is apparent that at a wetting transition, $\widetilde\beta=\pi/2$ and $\widetilde\alpha=\widetilde\gamma=0$, i.e.\ the tricuspid vanishes leaving only the $\alpha\gamma$ trajectory formed from the separate $\alpha\beta$ and $\beta\gamma$ paths, which meet at $90^\circ$, as required by the local scattering angle.
The wetting phase boundary itself follows by requiring that the trajectory touches the $\beta$ vertex at $x=\ell\cos\phi$, $y=\ell\sin\phi$, yielding
\begin{equation}
\frac{a}{\ell}\;=\;\sqrt{\;1+\frac{2}{\sqrt{3}}\,\sin\phi\;}
\label{wet}
\end{equation}
which determines all phase boundaries for wetting by $\alpha$, $\beta$ or $\gamma$.
This is the polar version of the wetting phase boundary which was obtained by Indekeu and Koga from Antonov's rule ($\sigma_{\alpha\gamma}=\sigma_{\alpha\beta}+\sigma_{\beta\gamma}$) using conjectures for the surface tensions.
For example, for the line of symmetry in the phase diagram where the bulk vertices form an isosceles triangle (obtained setting $\phi=\pi/2$ and $\ell=a^2$), we obtain $a=\sqrt{2\sqrt{3}-3}$, which was the original conjecture of Koga and Widom \cite{Koga2008,Koga2016}.
The analytical expression for the wetting phase boundary (\ref{wet}) also proves that CPW is absent in the KWI model along sections of the line of critical end points. More specifically, there is no CPW when the temperature scaling variable $t$ in the phase diagram lies between $t_\pm=(7\pm\sqrt{33})/8$, as conjectured by Indekeu and Koga \cite{Indekeu2022}.

In general, the trajectory is a quartic algebraic curve (i.e. degree $d=4$) possessing two singularities (i.e. $s=2$ corresponding to two double points, where $\partial_x u=\partial_y u=0$, situated at the $\alpha$ and $\gamma$ vertices).
The Riemann-Roch theorem for the genus, $g=(d-1)(d-2)/2-s$, then implies  $g=1$, so that, in general, the curve is elliptical, i.e.\ it can be parameterised on a rhombus (the fundamental domain of a torus) by two elliptic functions. However, at the wetting transition $g=0$, since additionally $\partial_x u=\partial_y u=0$ at the $\beta$ vertex, implying that $s=3$ and the curve may be rationally parameterised, i.e.\ it may be expressed as a quotient of polynomials in some parameter  - usually written as $t$ but not to be confused with the temperature scaling field mentioned above. For example, at the wetting transition occurring along the line of symmetry, we can write $(x(t),\tau\,x(t)-a^2)$, where $\tau=4b^2t/(t^2+b^2)$ with $b=\sqrt{2\sqrt{3}+3}$, $c^2=1\!+\!1/\sqrt{3}$ and
\begin{equation}
x(t)=\frac{4a^2\,\tau(3-\tau^2)(t^2+4b^2)\pm6\,c\,(\tau^2-a^2)\,(t^2-4b^2)}{3\,(\tau^4-6\tau^2+1)(t^2+4b^2)}
\label{rational}
\end{equation}
where the $\pm$ generate the $\alpha\beta$ and $\beta\gamma$ paths, respectively. 

Returning to the general properties of the trajectories, we note that the algebraic curve $u(x,y)$ is a harmonic function, satisfying the Laplace equation
\begin{equation}
\nabla^2u(x,y)=0
\label{Laplace}
\end{equation}
implying that the paths are conformally invariant – a property we may have suspected since 
the path equation (\ref{curve}) is homogeneous and, hence, invariant under any isotropic rescaling $x\to x/\rho$ and $y\to y/\rho$.
The conformal invariance applies throughout the \emph{whole} phase diagram and arises, not only from the isotropy of the KWI model but also because, in the mechanical analogy, the trajectories correspond to energy $\,E=0$.
With this observation we can now use the  methods of complex analysis and first determine the corresponding harmonic conjugate $v(x,y)$ from the Cauchy-Riemann equations, $\partial_x u=\partial_y v$ and $\partial_y u= -\partial_x v$. The harmonic conjugate, which also satisfies $\,\nabla^2 v=0\,$, is given by
\begin{equation}
    \begin{split}
        v(x,y)=  \cos\phi \left(\frac{3}{2}\,x^2y^2-\frac{x^4+y^4}{4}+\frac{\,a^2}{2}\,(x^2-y^2)\right)\\[.2cm]
          +\ell x \left(\frac{x^2}{3}-y^2-a^2\right)+\sin\phi\;\left(xy^3-x^3y+a^2xy\right)
    \end{split}
\label{v}
\end{equation}
and defines a family of curves that are orthogonal to $u(x,y)$.
The values of $v(x,y)$ at the ends of the trajectory are $v_\alpha=a^4\cos\phi/4+2\,\ell a^3/3$ and $v_\gamma=a^4\cos\phi/4-2\,\ell a^3/3$.
We note that, in the mechanical analogy, both $E=0$ and ${\bf p}\times\nabla v=0$ (with $\bf p$, the linear momentum) are conserved quantities.

With the harmonic functions $u(x,y)$ and $v(x,y)$, we can now define an analytic function $f(z)=u(x,y)+iv(x,y)$, of a single complex variable $z=x+iy$ given by
\begin{equation}
f(z)=-i\,e^{-i\phi} \left(\frac{z^4}{4} - \frac{a^2}{2} z^2 -\ell e^{i\phi}\left( \frac{z^3}{3} -a^2z \right)\right)
\label{f}
\end{equation}
which is a quartic complex polynomial.
The $\alpha\gamma$ trajectory and the two conjugate curves that pass through its end points are then just the real and imaginary parts of $f(z)=u_{\alpha\gamma}+iv_\alpha$ and $f(z)=u_{\alpha\gamma}+iv_\gamma$. Notice that the derivative of the function is $f'(z)=-ie^{-i\phi}(z-z_\alpha)(z-z_\beta)(z-z_\gamma)$, where $z_\alpha=-a$, $z_\beta=\ell e^{i\phi}$ and $z_\gamma=a$ are the bulk vertices in the complex plane.

To determine the $\alpha\gamma$ surface tension, we now exploit the conformal invariance. Any analytic function may be used to transform the trajectory into new coordinates but, by using the function $f(z)$ itself, we arrive at a remarkable simplicity. In this case, the conformal map $f(x+iy)=u+iv$ projects the $\alpha\gamma$ trajectory onto the vertical straight line $u=u_{\alpha\gamma}$ and similarly projects the two orthogonal curves onto the horizontal straight lines $v=v_\alpha$ and $v=v_\gamma$. This mapping is particularly revealing at a wetting transition since the conformal map not only flattens the $\alpha\gamma$ trajectory, but also maps the $\beta$ vertex onto the same line, i.e.\ it simultaneously maps the $\alpha\gamma$, $\alpha\beta$ and $\beta\gamma$ trajectories onto the same straight line  $u=u_{\alpha\gamma}$. This means that the lengths of the flattened trajectories are proportional to the surface tensions since the addition of their lengths is equivalent to Antonov’s rule, $\sigma_{\alpha\gamma}=\sigma_{\alpha\beta}+\sigma_{\beta\gamma}$. To prove this, we rewrite the expression for the surface tension (\ref{tension1}) as an integral along the $\alpha\gamma$ path in the complex plane. This gives $\sigma_{\alpha\gamma}=\sqrt{2}\int\!dz\, |f'(z)|$, since $|f'(z)|=|z-z_\alpha||z-z_\beta||z-z_\gamma|$,
equivalent to $|f'(z)|=\sqrt{\omega(x,y)}$. However, the integrand is just the local scale factor of the conformal map $f(z)$ that flattens the curve and, hence, $\sigma_{\alpha\gamma}=\sqrt{2}\,(v_\alpha -v_\gamma)$. Substituting $v_\alpha$ and $v_\gamma$ we obtain finally
\begin{equation}
\sigma_{\alpha\gamma}\;=\;\frac{4\sqrt{2}}{3}\;a^3\ell
\end{equation}
which is, precisely the Koga-Indekeu conjecture for the surface tension \cite{Koga2019}. This applies to all three interfaces, using the values of $a$ and $\ell$ defined for that side of the triangle, provided it is not completely wet by the other phase.
The conformal mapping ensures there is perfect consistency for the wetting phase boundaries as determined by the trajectory touching the $\beta$ vertex, or by using Antonov’s rule for the surface tensions. The proof of the surface tension formula also confirms that the wetting transitions are second-order. 

\begin{figure*}[t]
\includegraphics[width=1.8\columnwidth]{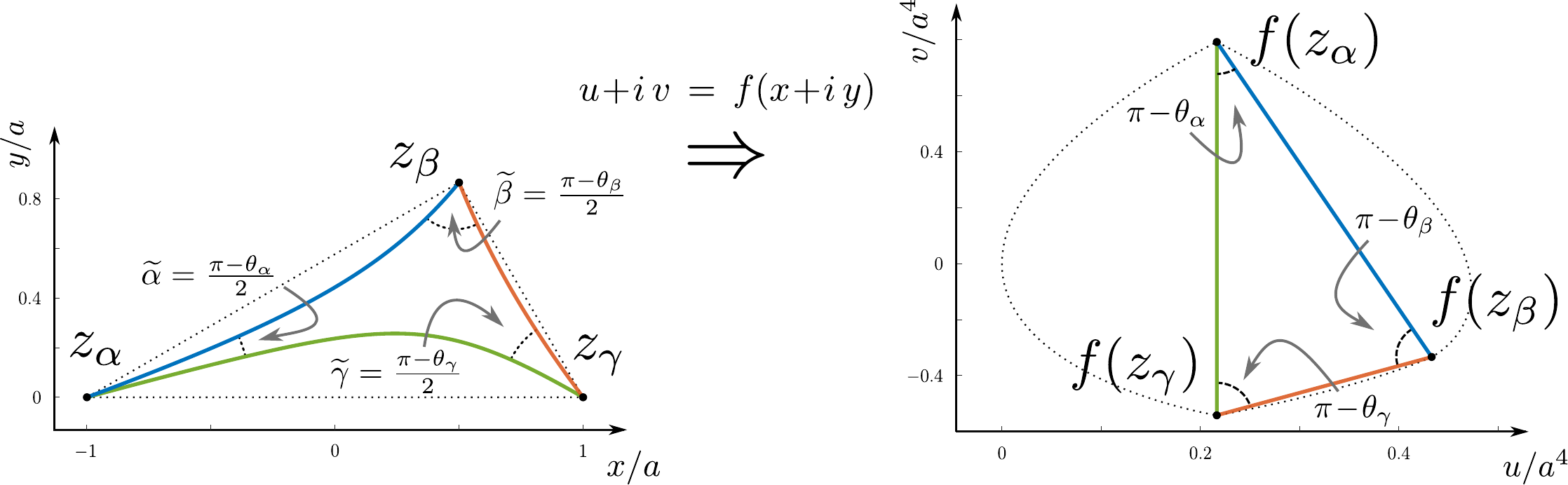}
\caption{\label{Fig3} 
Conformal mapping $f(x+iy)=u+iv$ of the three profile paths onto straight lines whose lengths are proportional to the corresponding surface tension, showing that the tricuspid is a curvilinear representation of the Neumann triangle. The transformation shows that the internal angles satisfy $\,\widetilde\alpha=(\pi-\theta_\alpha)/2$, and similarly for $\widetilde\beta$ and $\widetilde\gamma$.
The triangle connecting the bulk densities is also mapped (dotted lines).}
\end{figure*}

We can now appreciate the full significance of the tricuspid shape. The function $f(z)$ is the same, up to a rotation of an angle $\phi$, for all three paths. Therefore, the conformal transformation maps the tricuspid onto a triangle whose side lengths are proportional to the associated surface tensions, i.e.\ the tricuspid is a curvilinear representation of the Neumann triangle -- see Fig.\ \ref{Fig3}. The vertices of the triangle $z_\mu$ sit at the singularities where the conformal map fails (since $f'(z_\mu)=0$) and the angles are not preserved. However, near these singularities $f(z)\propto (z-z_\mu)^2$ and a simple application of de Moivre's theorem implies that the tricuspid angles are doubled under the mapping, and hence satisfy
\begin{equation}
\widetilde\alpha\,=\,\frac{\pi-\theta_\alpha}{2},\hspace{0.5cm}\widetilde\beta\,=\,\frac{\pi-\theta_\beta}{2},\hspace{0.5cm}\widetilde\gamma\,=\,\frac{\pi-\theta_\gamma}{2}
\label{neumann}
\end{equation}
which may also be verified by direct calculation. The remarkable implication of this is that the contact angles are encoded completely in the asymptotic decays of the density profiles, providing an unexpected link between global thermodynamic and entirely local quantities.  

To conclude, the KWI model is a very rare example of a density functional description of wetting which is analytically solvable – the single component square gradient theory \cite{Cahn1977,Nakanishi1982,Rowlinson1982} and the related Sullivan model \cite{Sullivan1979} being the other notable examples. The exact solution highlights some very surprising features - the underlying conformal invariance of the trajectories, and the link between the tricuspid shape and the Neumann triangle are particularly striking. Our analysis shows conclusively that Cahn's original speculation of the necessity of CPW is incorrect and partial wetting may persist up to the lines of critical end points, where the interfaces become infinitely diffuse due to the macroscopic size of the bulk correlation length. It is quite likely that this, and not CPW, is the more general scenario for wetting in fluid mixtures. Indeed, the prediction of a nonwetting gap along a section of the lines of critical end points, is already consistent with the experimental results reported in \cite{Kahlweit1989}. Further experiments may directly look for the wetting transitions, occurring close to the lines of critical end points, which are predicted by the KWI model.

\acknowledgments
We are deeply indebted to J.O.\ Indekeu and K.\ Koga for extensive correspondence and to R.P.W.\ Thomas, T.\ Bertrand, Y.\ Mart\'inez-Rat\'on and A.\ Malijevsk\'y for very helpful discussions. CR acknowledges the support  of grant PID2021-126307NB-C21(MCIN/AEI/10.13039/501100011033/FEDER,UE).

\bibliography{wetting}

\pagebreak

\end{document}